# Identification and Validation of the SNV Biomarkers Based on Multi-Dimensional Patterns

Boli, Lin Gao, Junying Zhang*, Liang Yu*

School of Computer Science and Technology, Xidian University, Xi'an 710071, Shaanxi, China

* To whom correspondence should be addressed: Email: lyu@xidian.edu.cn. Correspondence may also be addressed to Junying Zhang. Email: jyzhang@mail.xidian.edu.cn.

## Abstract

**Background:** Single nucleotide variants (SNVs) are detected as different distributions of DNA samples of distinct types of cancer patients. Even though, it is an exacting task to select the appropriate method to identify cancer to the greatest extent of SNVs.

**Results:** In this paper, we proposed a biomarker concept based on SNV patterns in different feature dimensions. Raw dataset (2761 samples) consisting of twelve different cancers was obtained from TCGA (The Cancer Genome Atlas). After preliminary screening of 562,321 DNA mutation sites in the samples, the mutation sites were extracted and characterized by cancer types in six different SNV feature dimensions. In this study, we found that the extracted features showed similar distribution in the cluster center of the disease type of the samples. After the initial processing of the raw data, the sample was more focused on the subtype distribution of the cancer or the cancer at the SNV level. We used k-nearest neighbors (KNN) to classify the extracted features and Leave-One-Out cross verified them. The accuracy of classifying is stable at around 97% and reached 97.43% at the highest. During the validation phase, we found validated oncogenes in the loci of the features with the highest importance among nine cancers.

**Conclusions:** In summary, the samples showed consistent patterns according to the cancer in which it belongs. It is feasible to classify the cancer of the sample by the distribution of different dimensions of the SNVs and has a high accuracy. And has potential implications for the discovery of cancer-causing genes.

## Introduction

Cancer is expected to rank as the leading cause of death and the single most important barrier to increasing life expectancy in every country of the world in the 21st century.[1] Large-scale tumour-sequencing studies have demonstrated that the majority of cancers are driven by either SNVs or SVs. There is currently a bias towards clinical sequencing of SNVs rather than SVs. [2]How to use SNVs to detect cancers as biomarkers efficiently and accurately is crucial. In earlier studies, using genetic mutations for cancer classification, the Cancer Genome Atlas Research Network performed mutational signature analysis on the core set of 196 hepatocellular carcinoma (HCC) applying a Bayesian variant of the non-negative matrix factorization (NMF) algorithm to mutation counts of SNVs stratified by 96-trinucleotide contexts.[3] In order to obtain the characteristics of the cancer, decompose the matrix used to carry the DNA mutation sites of the cancer by assuming that the DNA mutations is subjected to a specific distribution. Another earlier study developed a panel of DNA methylation biomarkers and validated their diagnostic efficiency for non-small cell lung cancer (NSCLC) in a large Chinese Han NSCLC retrospective cohort. The accuracy was 91% in their Bayes tree model.[4] These are the current two representative genetic mutation-based studies on cancer classification. The former research did not use specific genes as a basis, but uses trinucleotide to count the mutations of patients and identify the characteristics. The latter selected specific genes and used mutations in these genes to detect cancers. Due to the limitations of the data itself, the number of samples for these two studies is generally less than two hundred.

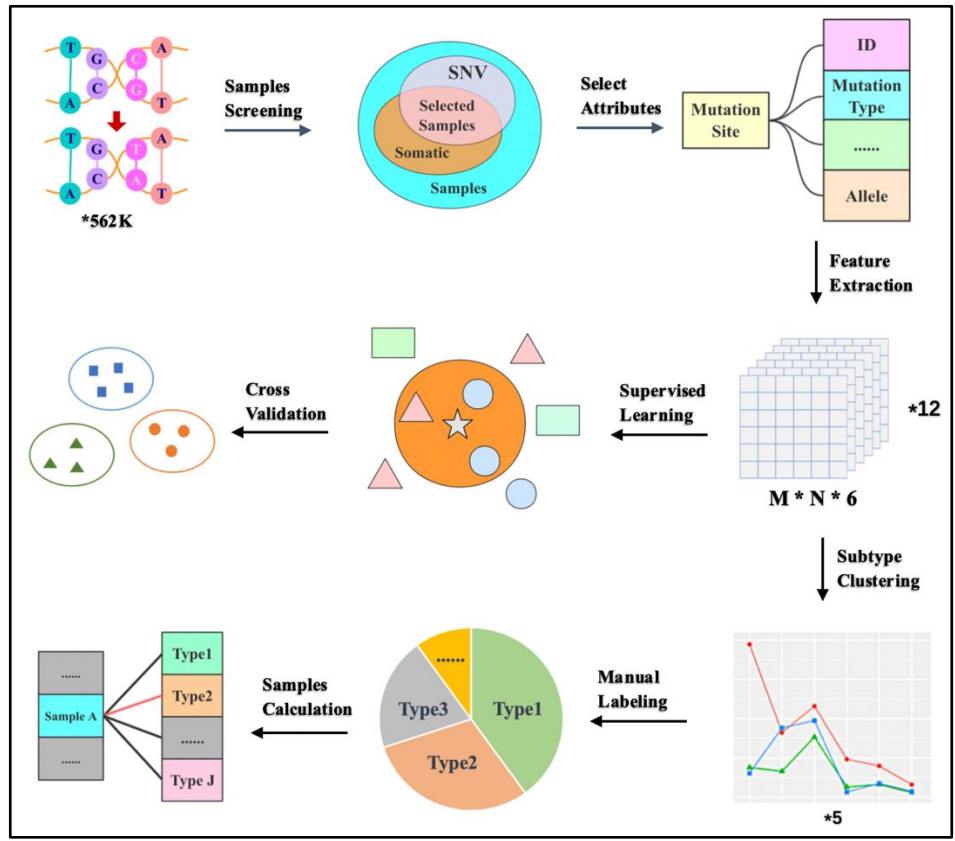

**Figure 1 The design framework of the study** Raw dataset contains 562K mutation sites which belong to 2761 samples. We screened all samples, and the retained samples must belong to SNV in somatic cells. And extracted different feature matrices as processed data with different feature dimensions. For the twelve cancers in the dataset, each cancer produced six M * N matrices, where M represents the number of samples contained in the cancer, and N represents the number of mutation types in the current feature pattern. Two operations were performed on the matrix: 1. The cancer subtypes are obtained by clustering the samples in the matrix. The samples in the matrix were manually labeled with subtypes and classified using Euclidean distance to verify the concept of SNV subtype distribution. 2. KNN classification was performed on the samples to verify the existence of specific distribution patterns separated by cancer in SNVs.

In this paper, to investigate the potential relationship between cancers and SNV patterns (Figure 1), We counted the SNVs of all patients with all twelve cancers, and extracted the features for each cancer according to six different feature dimensional extraction methods, then set up six mutation counting matrices to store six different patterns of mutation features. It was equivalent to expanding the dataset by six times. We classified all twelve cancers' samples and tested it using leave-one-out cross validation. The accuracy of cancer classification (KNN) by statistical SNV distribution patterns was 97.43%.

## Result

**Raw dataset collection, feature extraction and matrix construction**

We downloaded the eligible Mutation Annotation Format (MAF) files from TCGA as much as possible, which contained a large number of information about the mutation sites of twelve cancers, including their reference alleles and six to ten adjacent bases (both left and right), and mutant alleles at these sites. Twelve types of cancer include: Liver hepatocellular carcinoma (LIHC), Kidney renal papillary cell carcinoma (KIRP), Lung adenocarcinoma (LUAD), Prostate adenocarcinoma (PRAD), Lung squamous cell carcinoma (LUSC), Uterine Corpus Endometrial Carcinoma (UCEC), Thyroid carcinoma (THCA), Rectum adenocarcinoma (READ), Head and Neck squamous cell carcinoma (HNSC), Bladder Urothelial Carcinoma (BLCA), Acute Myeloid Leukemia (LAML), Pancreatic adenocarcinoma (PAAD).

The data obtained from TCGA cannot be used directly. Take LUSC as an example, it's MAF file contains 48,746 mutation sites belonging to 178 patients. The distribution of variants is shown in Figure 2. Cancer is a complex disease often associated with a characteristic series of somatic genetic variants.[5] Therefore, we only used somatic mutation sites where the mutation type is SNP (base substitution). After screening, 47,916 mutation sites were left.

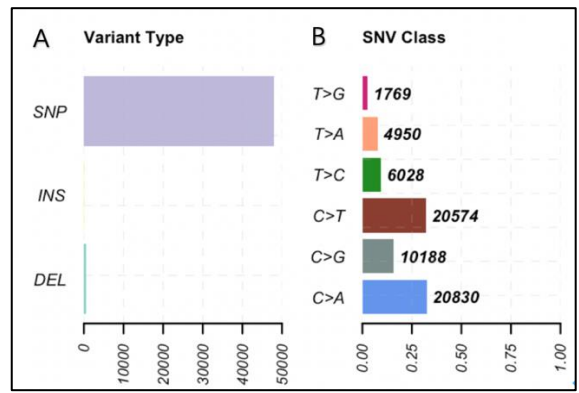

**Figure 2 Distribution of variants in LUSC.** **A** Distribution of SNV classes, the total number of mutation sites is 48,746, respectively belonging to SNP (substitution), INS (insertion), DEL (deletion). **B** Distribution of six basic base substitutions

The screening of mutation sites is the premise of feature extraction. After this step, we need to extract the features of interest from the original data, which are SNVs. We proposed the concept of the SNV feature dimensions to improve data utilization. The general classification of single nucleotide variations is based on the six kinds of base substitutions of nucleotides. For any SNV, the base (A, T, C, G) can be replaced by the remaining three, which result in 12 substitutions. Due to the complementary relationship of bases, the substitution is reduced to six (C>A, C>G,

C>T, T>A, T>C, T >G). In order to facilitate the distinction, we refer to such a one-dimensional feature by replacing one nucleotide of the mutation site itself.

The probability of a somatic single-nucleotide variant (SNV) to occur can depend on the sequence neighborhood[6]. Therefore, we combined the mutation site with its neighboring bases to form multi-dimensional features. For example, for a mutation site C>A, assuming that it's left nucleotides are T.C and the right nucleotides are G.A, obviously when taking a one-dimensional feature, only the mutation site itself is taken. The feature is C>A itself. When the feature dimension is extended to two-dimensional, two different features can be obtained due to different selection of the direction of feature expansion (when expanding to the left: C.C>A, when expanding to the right: C>A.G). The two-dimensional features may add a single base to the left or right to make the possible mutation type 2*24. Similarly, when the feature dimension is extended to three-dimensional, we can again acquire three new sets of features (when expanding to the left: T.C.C>A, when expanding to both sides: C.C>A.G , when expanding to the right: C>A.G.A). Among these three sets of features, each set of features has 96 types that may appear. Considering that the distribution of data has been sparse when expanding to three-dimensional, we only extended the features to three-dimensional, which gave us six sets of features.

After the feature extraction patterns is established, the sample features can be statistically calculated. We converted the sample features into six groups of M*N matrices in six different patterns (M is the number of samples, and N is the number of mutation types in the current pattern). For example, with regard to 178 patients with LUSC, using one-dimensional as the feature extraction pattern, a 178*6 matrix W1 can be established. W1's any element $W_{ij}$ represents number of No. j mutation in the No. i sample. By traversing the 47,916 mutation sites in turn, we can obtain a complete one-dimensional feature matrix. The sum of all elements of the matrix is equal to the total number of mutation sites. After the feature collection of the data in the order of the six feature patterns, the different patterns are equivalent to observing the data at different angles, which is equivalent to expanding the data by six times. Once these matrices are obtained, the processing of the data is complete, then we can analyze the association of the data with cancers.

**Association of SNV biomarkers with cancer subtypes**

In our earlier work, we selected PRAD, LIHC, and LUAD as the initial dataset, and converted the distribution of all SNVs in the same cancer into six groups of matrices representing the distribution of various types of mutation sites. In order to confirm the existence of a special SNV overall distribution for each cancer, we added the performance of each row in the matrix by the mutation type of each pattern and divided by the total number of samples to obtain a uniform distribution of the samples. When we tried to classify the test set by using the obtained uniform distribution as the standard, we found that the effect was not stable. The performance of PRAD was satisfied with expectations, the classification accuracy was about 98%, but LIHC and LUAD could only reach 70%~80%.

The prevalence of somatic mutations was highly variable between and within cancer classes.[7] We have observed all the samples individually to find out the problem. To control experiment cycle and cost, we selected five cancers from the raw data set to form a sub-dataset, including Lung Adenocarcinoma (LUAD), Lung squamous cell carcinoma (LUSC), Prostate adenocarcinoma (PRAD), Kidney renal papillary cell carcinoma (KIRP), Liver hepatocellular carcinoma (LIHC). In this way, the diseased areas of cancer were differentiated as a whole, yet the association between different cancers in the same diseased area (LUAD and LUSC) was also considered. Study showed that even with the same cancer, the internal SNV distribution does not necessarily obey only one single overall distribution. However, the occurrence of SNV is not completely random. In the five cancers of the dataset, SNVs are often subjected to one to three major distributions. The main distributions of the five cancers is shown in Figure 3.

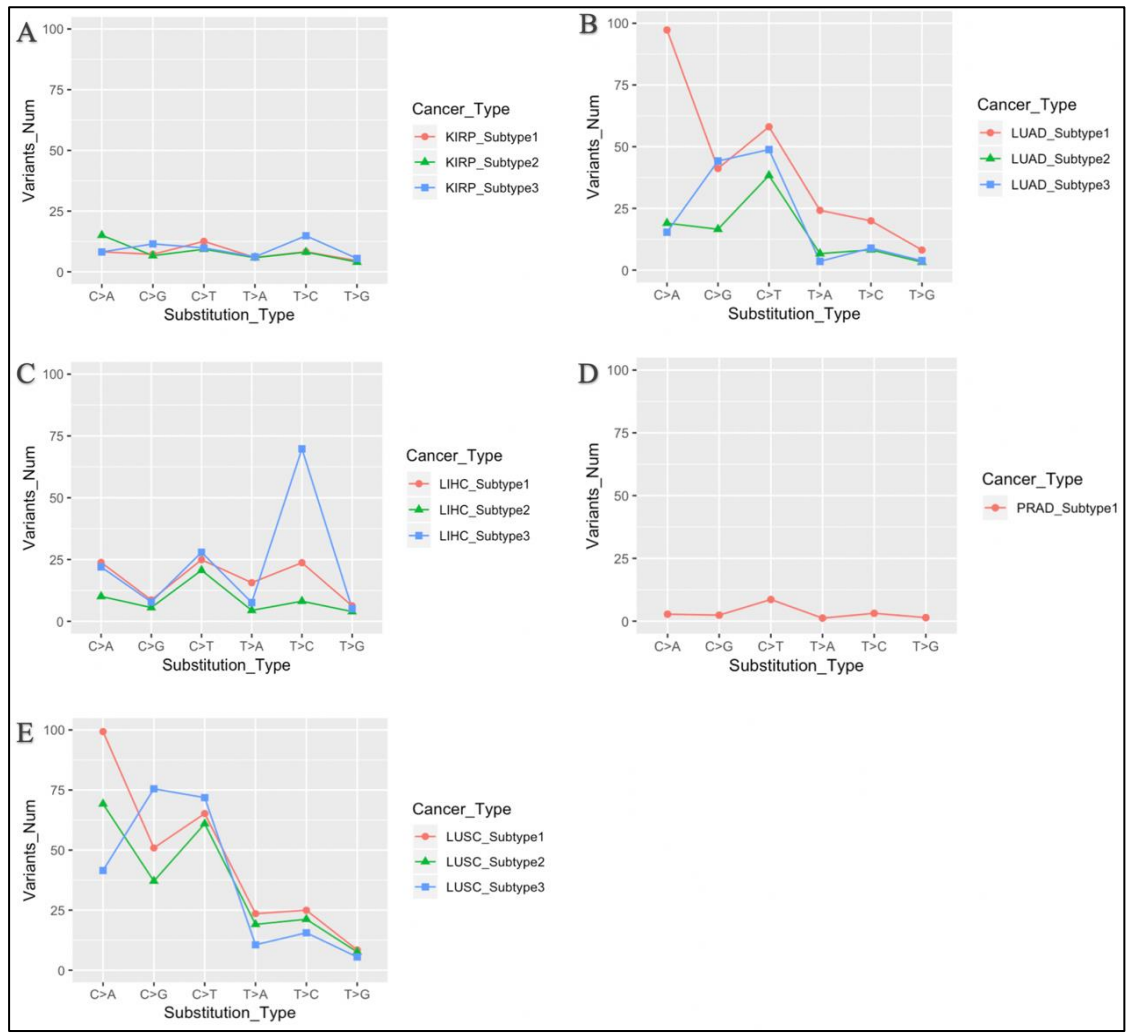

**Figure 3 One-dimensional feature distribution of cancers** Since high-dimensional features are extensions of low-dimensional features, one-dimensional features were used here as representative of the distribution of SNVs in cancers. Each line chart illustrates the distributions of major subtypes in the belonging cancer. The abscissa represents the six types of mutation site substitutions in the one-dimensional pattern, and the ordinate represents the average number of each substitution type of all samples that belong to the subtype in the current cancer.

It is worth noticing that the labeled subtypes do not fully cover all the samples. This is because the subtype labels were manually labeled by the researcher. Before we labeled the subtype labels, we first calculated the mean distribution of all the samples' one-dimensional features, and this distribution was referred to as the first subtype. The samples that apparently did not conform to the first subtype were then manually selected as new subtypes. The number of samples selected for the new subtype must be greater than 10%, which means, it is not accidental, but there is indeed a portion of the samples around this distribution. We selected three subtypes for each type of cancer at most to avoid overfitting, and samples that did not conform to the selected subtypes remained in the first subtype cluster.

To verify that these subtypes do exist, we randomly used 80% of the dataset as the training set and 20% of the dataset as the test set. In order to clearly obtain the classification results, we selectively filtered some samples that should satisfy: very inconsistent with any established subtypes, the mutations that occur are too few or too large to be analyzed, too close to other cancer subtypes will inevitably be classified incorrectly, and all three types must satisfy the condition that the number of occurrences is rarely enough to constitute a new subtype. The above sample filtering was only for the purpose of observing the classification results of subtypes, and has been removed in practical applications. Since the distribution of SNVs in the samples is expressed in digital form, we used different formulas to calculate the distances between the test set and the 13 subtypes in the training set in the six feature patterns, and linearly accumulate these distances. The Euclidean distance with the best classification result is selected as the calculation standard, we took the nearest subtype as the classification result. Compare to defaulting the SNVs of a cancer to consistent distribution, the classification results of subtypes have improved significantly and stably, from 80% to 91%. The distance calculation formula is as follows:

$$Dist(X,Y) = \sum_{i=1}^{6} \sqrt{\sum_{j=1}^{n_i} (x_{ij} - y_{ij})^2}$$

X and Y represent any two samples in the sample set, i represents six feature extraction patterns, $n_i$ represents the number of base substitution types in different feature patterns, j represents each type of base substitution, $x_{ij}$ and $y_{ij}$ represent the number of No. j substitution in the No. i pattern.

It should be noted that we found that the distance between the sample with the wrong classification and the subtype of the cancer that the sample actually belongs to is often very close to the distance of the misclassified cancer and the sample, which means that even the sample with the wrong classification is not uncontributed. We chose the subtype with the smallest distance as the classification result to obtain the classification result intuitively, but in practical application, we should consider the case where the distance is close, so that even the misclassified samples could narrow the classification result to very small range.

**Classification of cancer based on SNV biomarkers and validation**

In later studies, to actually apply the SNV multi-feature patterns to cancer classification, we used a better algorithm to classify the samples instead of simply performing distance estimation. K-Nearest Neighbor (KNN) is considered a very common algorithm when using distance to measure samples[8], and KNN is also very suitable for the data type of the samples. We used Euclidean distance that performed well in previous studies to calculate the distance between test samples and other labeled samples. The experimental results of KNN are shown in Table 1 and Figure 4.

**Table 1 Sample numbers and classification results of twelve cancers** We found that the accuracy of LAML is lower than other cancers. This is because there are several samples with too sparse mutations in LAML, which cannot make the samples well classified.

| Cancer Type | Sample Number | 2-NN Accuracy |
|---|---|---|
| LIHC | 198 | 0.9697 |
| KIRP | 282 | 0.9716 |
| LUAD | 230 | 0.9870 |
| PRAD | 259 | 0.9575 |
| LUSC | 178 | 0.9775 |
| UCEC | 194 | 0.9897 |
| THCA | 425 | 0.9765 |
| READ | 161 | 0.9938 |
| HNSC | 364 | 0.9780 |
| BLCA | 130 | 0.9846 |
| LAML | 151 | 0.9338 |
| PAAD | 189 | 0.9683 |
| ALL | 2761 | 0.9743 |

In order to verify our proposed marker concept, we ranked the genes included in the top features by the number of mutations by ranking the importance of 342 features in nine cancers. Discovered oncogenic genes have been found in every type of cancer, including NARS and HARS in THCA; KRAS in LUAD, etc. This not only proves the effectiveness of SNV as a marker, but also provides research ideas for potential oncogenes.

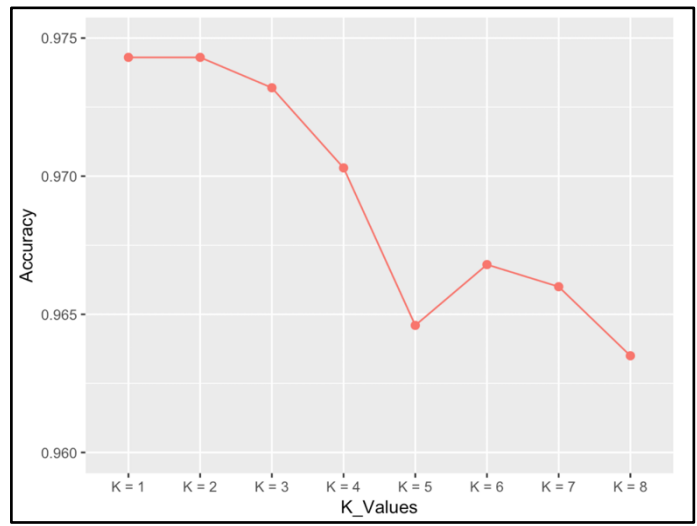

**Figure 4 The effect of K value on accuracy** We performed a classification effect test on different values of K. Performed Leave-One-Out cross validation for each K value in the 12 cancer classifications. The accuracy reached the highest (97.43%) when k was 1 or 2.

## Conclusions and future work

In this paper, we found that although SNVs are altered in cancer, they still follow one or several overall distributions. In the experiment, we demonstrated the positive effects of multi-dimensional features of SNV biomarkers on cancer classification, extending only to three-dimensional, the classification accuracy is improved by 10% to 20%. Efficient cancer classification was achieved by using the KNN algorithm, and the accuracy reached 97.43% at the highest.

In fact, because our proposed classification of cancer based on SNV patterns is a brand-new method, after investigation of other studies, we found that the study of multi-class cancer classification is generally to analyze images (x-rays) or genes. The data we used did not find that pre-study studies were used in multi-class cancer classification, since the datasets were different, there was no gold standard set, which made it difficult to compare the experiments. However, we found that the accuracy of existing research on multi-class cancer classification is generally 85% to 95% [9-12], which indirectly proves that our model is valuable.

Limited to the source of experimental data, we haven't verified the association between SNV subtypes and cancer subtypes. In the future work, we can further refine the subtypes of cancer and explore the relationship between SNV patterns and cancers. In addition, multi-class cancer classification using the multi-dimensional SNV patterns is also feasible and has high precision.

## Core code and example dataset

Our core code and example dataset have been written as a R package and can be found at:

https://github.com/Lee0510/SNVCC